\documentclass[aps,showpacs]{revtex4}
\usepackage{amsmath}
\usepackage{graphicx}
\def\ra{\rangle}
\def\la{\langle}

\newcommand{\vsig}{\vec{\sigma}}

\newcommand{\vSIG}{\vec{\Sigma}}

\def\lsim{\mathrel{\raise3pt\hbox to 8pt{\raise -6pt\hbox{$\sim$}\hss{$<$}}}}
\def\rsim{\mathrel{\raise3pt\hbox to 8pt{\raise -6pt\hbox{$\sim$}\hss{$>$}}}}

\def\lsim{\mathrel{\raise3pt\hbox to 8pt{\raise -6pt\hbox{$\sim$}\hss{$<$}}}}

\def\ra{\rangle}
\def\la{\langle}

\begin{document}
\bibliographystyle{apsrev}

\draft
\title{Weak Magnetism Correction to Allowed Beta-decay for Reactor Antineutrino Spectra}

\author{X.B. Wang}
\affiliation{School of Science, Huzhou University, Huzhou 313000, China}
\author{A.C. Hayes}
\affiliation{Los Alamos National Laboratory, Los Alamos, New Mexico 87545, USA}

\date{\today}

\begin{abstract}
The weak magnetism correction and its uncertainty to nuclear beta-decay play a major role in determining the significance
of the reactor neutrino anomaly. Here we examine the common approximation used for one-body 
weak magnetism in the calculation of fission 
antineutrino spectra, wherein matrix elements of the orbital angular momentum operator
contribution to the magnetic dipole current are assumed to be proportional to those of the spin operator.
Although we find this approximation invalid for a large set of nuclear structure situations, we conclude that it is valid for the relevant allowed beta-decays between fission fragments.  In particular, the uncertainty in the fission antineutrino due to the uncertainty in the one-body weak magnetism correction is found to be less than 1\%.  
Thus, the dominant uncertainty from weak magnetism for reactor neutrino fluxes lies in the uncertainty in the two-body meson-exchange magnetic dipole current.
\end{abstract}

\pacs{24.80.+y, 11.30.Er, 24.60.-k, 21.30.Fe}
	
\maketitle

\section{Introduction}
There has been considerable interest recently in the expected
 magnitude and shape of the fission aggregate antineutrino spectra emitted from reactors.
This interest has been driven by the fact that reevaluations \cite{mueller,huber} of the spectra 
for all four actinides ($^{235}$U, $^{238}$U, $^{239}$Pu and $^{241}$Pu) contributing to reactor antineutrino fluxes
led to a systematic increase in the expected
number of antineutrinos above about 2 MeV, relative to earlier evaluations \cite{Schreckenbach, vogel-1}, which, in turn, led
to the so-called reactor neutrino anomaly \cite{anomaly}.
The changes in the evaluated spectra were in part due to  changes in the treatment of sub-dominant corrections to nuclear beta decay, particularly the treatment of the finite-size and weak-magnetism corrections.

There are four sub-dominant corrections to beta decay that must be considered in calculating the antineutrino spectra.
These are the recoil, radiative, finite-size, and weak-magnetism corrections. 
The recoil correction is quite small, and the radiative correction has been taken from the work of Sirlin \cite{sirlin, sirlin-new}, both before and since the occurrence of the anomaly.
In the earlier work of Schreckenbach {\it et al.} \cite{Schreckenbach}, the finite-size and weak-magnetism corrections were
applied to the aggregate fission 
antineutrino spectrum as a single energy-dependent correction to the entire spectrum, whereas in the
reevaluations \cite{mueller,huber} these corrections were applied for each end-point energy (or end-point energy range)
contributing to the spectrum. 
In all cases,  the finite-size and weak-magnetism corrections that were applied involve some level of approximation.
For example, the finite-size correction is often taken to be the form appropriate for allowed beta decay transitions 
although it is well recognized \cite{Hayes14,hayes2,bnl} that about 30\% of the 
transitions making up the antineutrino spectra are forbidden.
In the case of the 70\% allowed transitions, 
the expression \cite{Hayes14,holstein} for the finite-size correction has been found \cite{zemach16} to be reasonably accurate.
For the weak magnetism correction, there is both a  one-body and  two-body contribution.
As discussed in detail below, in case of the one-body weak-magnetism correction, both a 
spin $\vec{\Sigma}=\sum_i  \tau_i^{\pm}\, \vsig_i$ and 
orbital angular momentum  $\vec{\Lambda}=\sum_i  \tau_i^{\pm}\, l_i$ operator enter, 
 where $\vec{\sigma}_i=2\vec{S}_i$ and $\vec{l}_i=\vec{x}_i\times\vec{p}_i$ for nucleon $i$.
Thus, in general, the weak-magnetism correction is nuclear structure dependent.
However, in most analyses of reactor antineutrino spectra \cite{Hayes14, vogel-1}, 
the nuclear structure dependence is simplified by assuming that
matrix elements of the orbital angular momentum operator are  proportional to matrix elements of the spin operator, i.e.,
it is assumed that 
$\la J_f ||\vec{\Lambda} ||J_i \ra =-\frac{1}{2}\la J_f ||\vec{\Sigma} ||J_i \ra$.
The  purpose of the present work is to check the validity of the latter nuclear structure 
assumptioni, particularly for allowed fission fragment beta-decays.

\subsection{The Weak Magnetism correction}

In earlier work \cite{zemach16}, we  examined the usual approximations made for the finite-size correction to allowed beta decay 
by evaluating the Zemach moments using energy density functional theory, and found these approximations to be good ~\cite{zemach16}.
The approximations made for the weak-magnetism correction are of an entirely different origin, and need to be assessed separately.
Weak magnetism is generally a small correction to beta-decay, that arises to first order 
 from an interference term between the dominant Gamow-Teller contribution and the magnetic dipole contribution to the weak current.
The magnetic dipole operator,
$\vec{\mu}_V=\frac{1}{2}\int d^3x~\vec{x}\times\vec{J}_V(\vec{x})$, involves the sum of three terms in the vector current $\vec{J}_V(\vec{x})$, namely, the spin current, the orbital current, and meson-exchange currents.
\begin{equation}
\vec{\mu}_V=\vec{\mu}_s+\vec{\mu}_L+\vec{\mu}_{MEC}=\frac{\mu_v}{2M_N}\vec{\Sigma}+\frac{1}{2M_N}\vec{\Lambda}+\mu_{MEC}.
\label{mu-terms}
\end{equation}
The label ``MEC" indicates leading-order meson-exchange currents, which are dominated by pion-exchange, and whose form are
not a subject of the current work.
We note that, unless matrix elements of $\vec{\Sigma}$ are suppressed, 
the first term in eq.(\ref{mu-terms}) tends to dominate because the vector magnetic moment is large compared to unity, i.e., 
$\mu_v=4.7$.

The first order contribution from the  weak magnetism correction $\delta_{WM}$
  to the beta-decay spectrum for allowed and and first forbidden decays
is given in ref. \cite{Hayes14}, and for allowed GT transitions is,
\begin{eqnarray}
\frac{\!\!d\omega}{dE_e} &=& \frac{G_F^2\, \cos^2{\theta_C}}{2\pi^3}\, p_e\, E_e (E_0-E_e)^2 F(e,Z)   g_A^2  \mid\la\vSIG \ra\mid^2(1+\delta_{WM}) \\
&=&\frac{G_F^2\, \cos^2{\theta_C}}{2\pi^3}\, p_e\, E_e (E_0-E_e)^2 F(e,Z)   g_A^2  \mid\la\vSIG \ra\mid^2\left(1+ \frac{4}{3}\left[\frac{\mu_v+\frac{\la J_f\mid\mid\vec{\Lambda}\mid\mid J_i\ra}{\la J_f\mid\mid\vec{\Sigma}\mid\mid J_i\ra}}{2\,M_N\,g_a}\right](2 E_e -m_e^2/E_e-E_0)\right).
\label{WM}
\end{eqnarray}
where $G_F (g_A)$ is the Fermi (axial) coupling constant, $F(E,Z)$ is the relativistic point Fermi function, $E_e (m_e)$ are the electron energy and mass, and $E_0$ is the end-point energy for the transition. 
For the sake of brevity
we do not list other sub-dominant corrections such as recoil, radiative, and finite size corrections; they are discussed in the
review \cite{hayes-vogel}.

In several references \cite{vogel-1,mueller,huber,Hayes14}, a simple model was used in which the orbital angular momentum of the $ith$ nucleon $\vec{\ell}_i$ was replaced by $\vec{j}_i-\vec{\sigma}_i/2 $, and it was argued that any change in the quantum numbers of the $\beta$-decaying nucleon would eliminate
 the $\vec{j}_i$ term, so that this term could be dropped. This simple model effectively assumes that 
\begin{equation}
\delta_{LS}^{j_fj_i}\equiv \frac{\la J_f\mid\mid\vec{\Lambda}\mid\mid J_i\ra}{\la J_f\mid\mid\vec{\Sigma}\mid\mid J_i\ra} \simeq -\frac{1}{2}
\label{bigeq}
\end{equation}
in eq.(\ref{WM}).
Examining the validity of this approximation, which replaces the fractional contribution from
the  orbital currents by ``-1/2" in the expression for weak-magnetism correction, is the central focus of the present work.
In addition, we examine whether matrix elements of 
the operator $\vec{\Sigma}$ truly dominate  over  those for the orbital angular momentum
 operator $\vec{\Lambda}$, for the set of transitions relevant to reactor antineutrino spectra.
\section{Nuclear Structure Studies of the Fission Fragment beta-decays}

\subsection{Single-particle matrix elements}
We begin with a discussion of the simple case of pure single-particle matrix elements (SPME) for the spin and orbital currents involved in the weak-magnetism correction.
In this case, the fractional orbital correction is,
\begin{eqnarray}\label{delta_ls_sp}
\delta_{LS} ^{j_fj_i}=&& \frac{\la n l j_f||\vec{\Lambda}  ||n l j_i\ra}{\la n l j_f\ ||\vec{\Sigma} ||n l j_i\ra}=(-1)^{J_i-J_f} \frac{\left\{\begin{array}{ccc}1/2 & l & j_i \\1 & j_f & l\end{array}\right\}}{\left\{\begin{array}{ccc}l & 1/2 & j_i \\1 & j_f & 1/2\end{array}\right\}}\frac{\sqrt{l(l+1)(2l+1)}}{\sqrt{6}}\,,
\end{eqnarray}
where ``$n l_i j_i$" and ``$n l_f j_f$" are the single particle orbits involved in the initial and final states, and $ l_i= l_f=l$ for the allowed transitions. 
We introduce the following short-hand notation: with  $j_i = l \mp 1/2$ and $j_f=l \pm 1/2$, we use the signs in these definitions to stand for $j_i$ and $j_f$ in $\delta_{LS} ^{j_fj_i}$ (Eq.~(\ref{bigeq})), {\it i.e.},  $\delta_{LS} ^{+-}$ stands for  $j_i = l - 1/2$ and $j_f=l + 1/2$, and {\it etc.}. 
From the analytical expressions for the $6-J$ symbols,  listed in Ref.~\cite{Edmonds60}, we find,
\begin{eqnarray}\label{delta+-}
\delta_{LS} ^{--}=-(l+1),&&\quad \delta_{LS} ^{-+}=-1/2, \nonumber\\
\delta_{LS} ^{+-}=-1/2,&&\quad \delta_{LS} ^{++}=+l.
\end{eqnarray}
Thus, if the overlap of the initial and final radial wave functions involved in the $\beta$-decay transition is close to unity,
replacing the fractional  orbital correction by  ``$-1/2$" can be a good approximation, when only one single-particle orbit from the initial and  final state are involved in the transition and  $j_i \neq j_f$ ($\delta_{LS} ^{-+}$ and $\delta_{LS} ^{+-}$).
This is particularly true for spin-orbit partners.

\subsection{Hartree-Fock-Bogoliubov calculations}
Moving beyond the case of simple single-particle transitions
 requires a model to describe the structure of the fission fragments whose beta-decays make up the cumulative fission antineutrino spectra.
In Fig.~1, we show the neutron and proton numbers for the dominant fission fragments contributing to the antineutrino spectra of
 $^{235}$U,  $^{238}$U, $^{239}$Pu, and $^{241}$Pu. Displayed is the set of fission fragments listed in the ENDF/B-VII.1 nuclear database
with cumulative fission yields $>\sim 0.01$, Q value $>\sim$ 3.0 MeV, half life less than 1 month, and those that are listed as decaying by allowed transitions. 
In all case, the large neutron excess means that the Fermi surfaces for neutrons and protons are quite different.
\begin{figure}
\includegraphics[angle=0,width=12 cm]{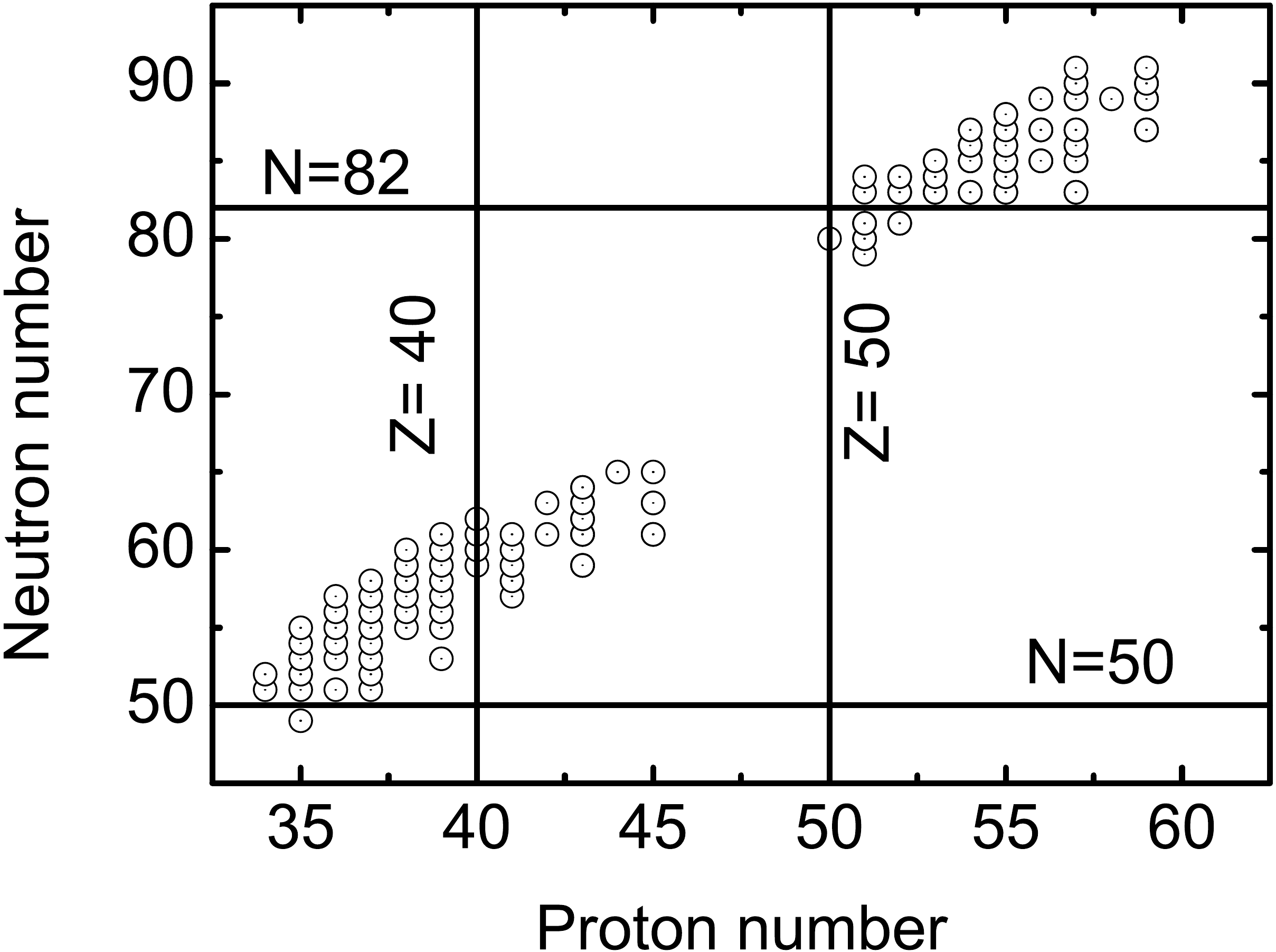}
\caption[T]{\label{fig:Fy}
The proton and neutron numbers for the dominant fission fragments contributing to the 
aggregate fission antineutrino spectra for   $^{235,238}$U and $^{239,241}$Pu. }
\end{figure}

For all of these nuclei, any nuclear structure calculation has to involve a model-space truncation.
To model these nuclei, we  ran a series of Hartree-Fock-Bogoliubov (HFB) calculations.
Compared to shell model calculations, 
HFB calculations can be realized in considerably larger model space. 
A main goal in the HBF study was to determine the effect of the pairing interaction on the nuclear wave functions, and the resulting effect on the ratio $\delta_{LS}^{j_fj_i}$. For nuclei close to the drip-line, 
the pairing interaction is found to scatter nucleon pairs from bound states to positive-energy orbitals, which in some cases leads to quenching of shell effects.   
For the HFB calculations, we used the Skyrme SLY4 parameterization~\cite{Cha1998} for the particle-hole interaction.
A standard zero-range density-dependent pairing force was used for the particle-particle channel, 
\begin{equation}
f(r)=V_0\left(1-0.5\frac{\rho}{\rho_0} \right),
\end{equation}
where $V_0$ is the interaction strength and $\rho_0$ was fixed at 0.16 fm$^{-3}$. 
We used the Lipkin-Nogami (LN) method \cite{lipkin} to restore particle number approximately. The numerical code {\sc HFODD}~\cite{hfodd} was used for these calculations. 
We adopted the commonly used equivalent-spectrum cutoff of 60 MeV, applied in the quasiparticle configuration space, and the calculations were performed in a spherical basis of 14 major harmonic-oscillator shells. 
The value of $V_0$ was -272.76 MeV fm$^3$, which gives an empirical neutron pairing gap for $^{120}$Sn of 1.245 MeV. 

In Fig.~2, we show the ``equivalent spectrum" of single-particle energies and the corresponding occupation coefficients,
 as defined in Ref.~\cite{NPA1984}, from our
HFB-LN calculations, for nuclei  in the ``$1g2d3s+1h11/2$" shell with $A<100$ 
Fig.~3 shows the same but for nuclei in the ``$1h2f3p$" shells with $A>130$.
 Only even-even nuclei that are contained within the set of fission fragments in Fig.~\ref{fig:Fy} are shown. 

\begin{figure}
\includegraphics[angle=0,width=12 cm]{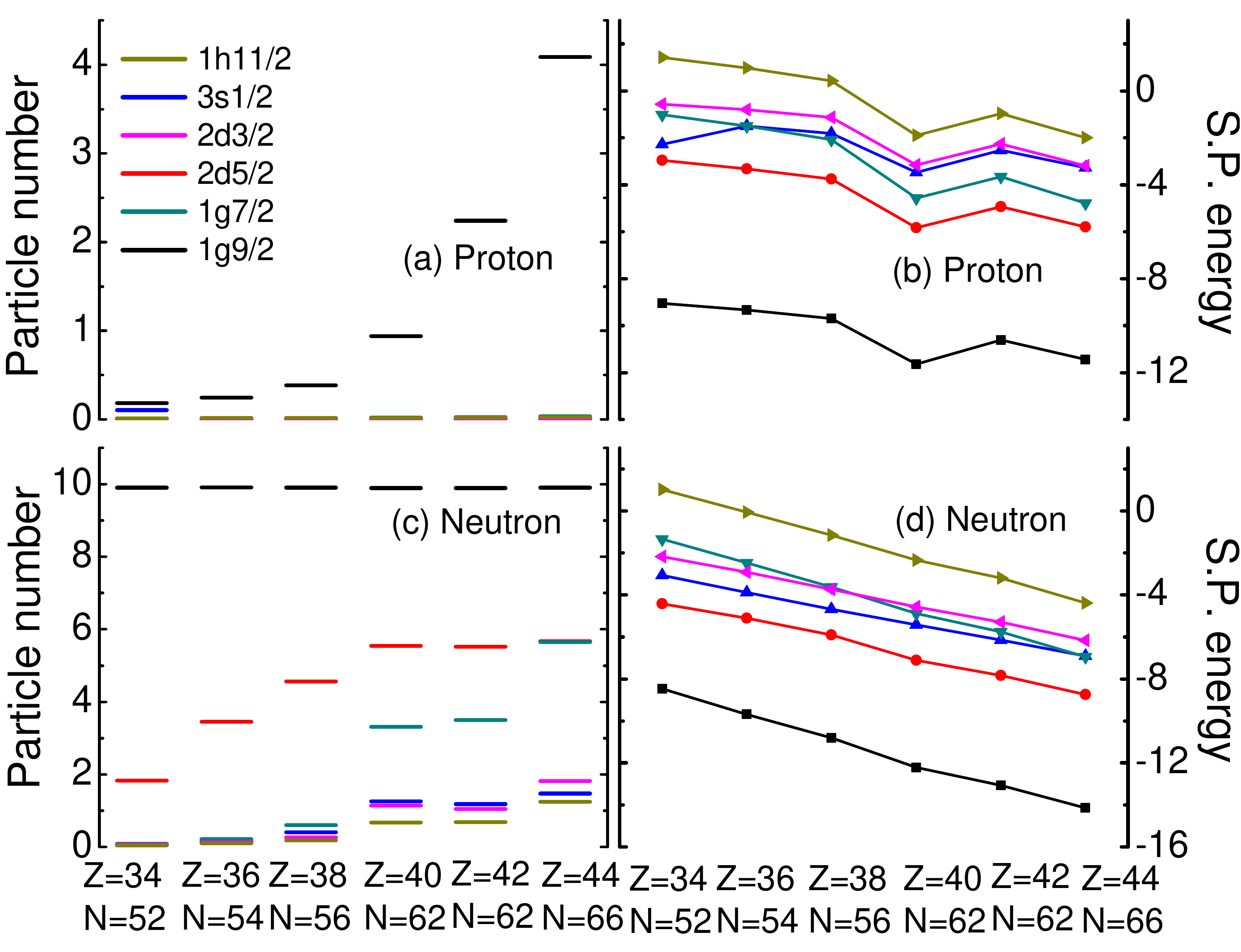}
\caption[T]{\label{fig:A100}
(Color online) The occupation probabilities (particle number) and ``equivalent-spectrum'' for the proton and neutron 
 single-particle energies, as predicted by out HFB-LN calculations for nuclei with $A \leq 110$ and in the ``$1g2d3s1h11/2$" shell.}
\end{figure}

\begin{figure}
\includegraphics[angle=0,width=12 cm]{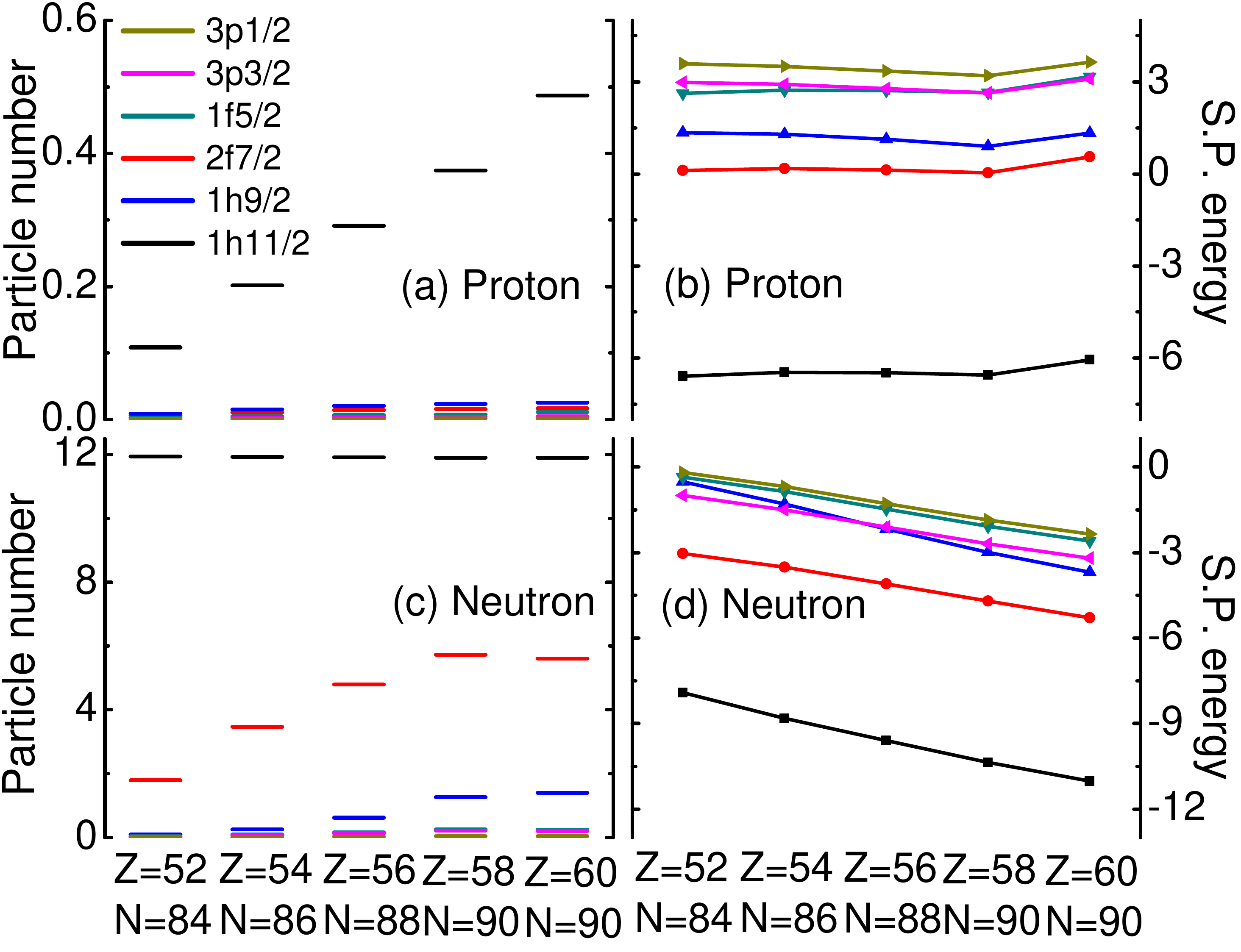}
\caption[T]{\label{fig:A130}
(Color online) The same as Fig.~\ref{fig:A100}, but for nuclei with $A>130$ and in the ``$1h2f3p$" shell.}
\end{figure}

In this calculation we have assumed spherical shapes and studied the shell gaps in these nuclei.
For the fission fragments of mass A  $\sim 100$, there is a large shell gap between the $1g9/2$ and $1g7/2$ orbitals, shown in Fig.~\ref{fig:A100}. When the nuclear interaction is turned on, $1g9/2$ valence protons remains in $1g9/2$ with large 
probability; even with the pairing interaction,  the occupation probabilities in orbitals higher than $1g9/2$ is negligible. 
For neutrons, in Fig.~\ref{fig:A100} (c), the $1g9/2$ shell is nearly filled, and the 
valence $1g7/2$ and $2d5/2$ orbitals play a major role. 
Thus, the allowed beta-decay transitions for the lower mass fission fragments 
are dominated by neutron transitions from
the $1g7/2$ orbital to the proton $1g9/2$ orbital.
For typical fission fragments of mass $A>130$, Fig.~\ref{fig:A130}, there is  a large shell gap between $1h11/2$ and $1h9/2$ orbitals, and the beta decays are dominated by  $1h9/2$  neutron transitions to the $1h11/2$ proton orbital.

\subsection{Shell-model calculations}
\begin{figure}
\includegraphics[angle=0,width=12 cm]{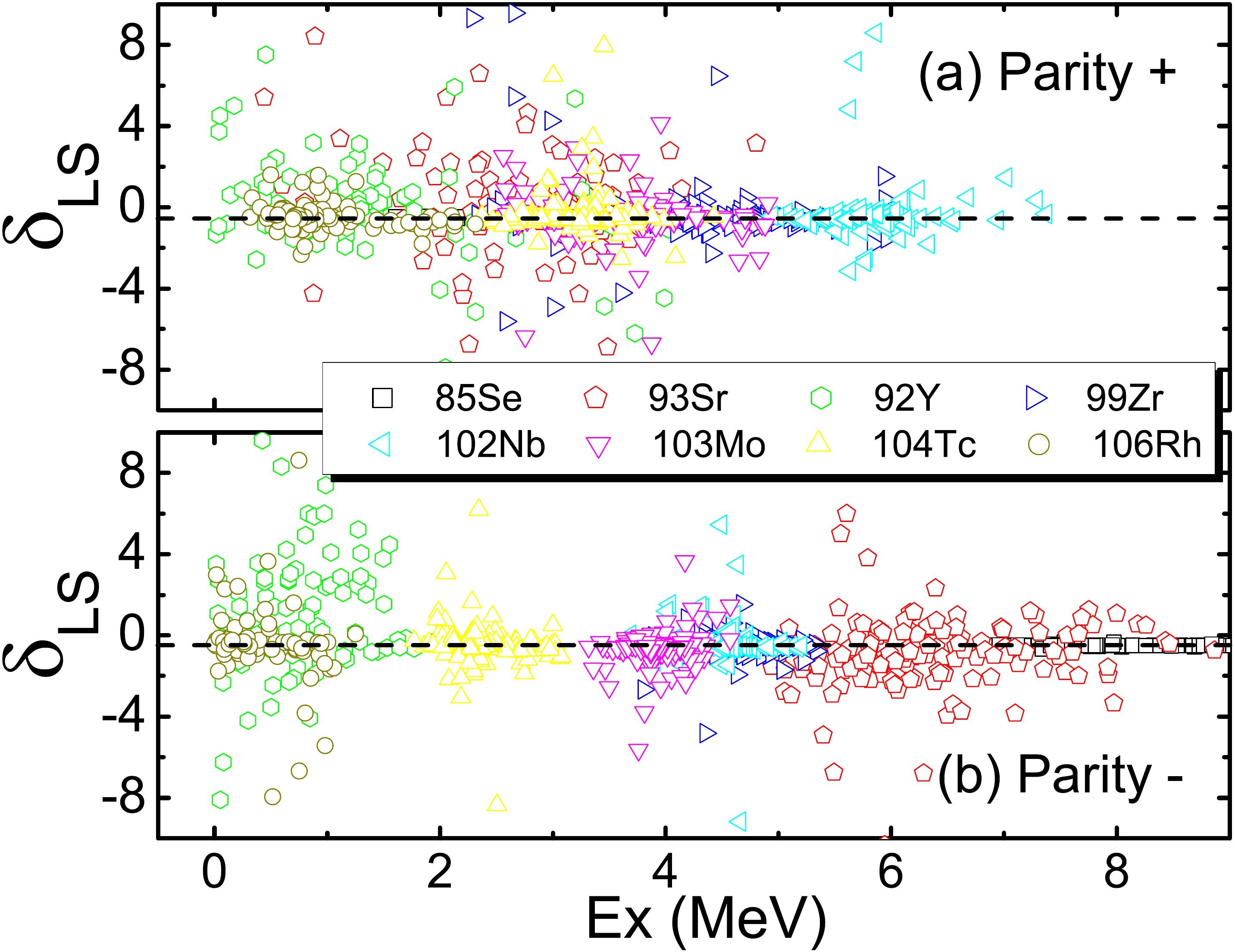}
\caption[T]{\label{fig:delLS}
(Color online)  The values of $\delta_{LS}^{j_fj_i}$ calculated for 
the allowed beta-transitions between the  lowest 20 positive parity (panel (a)) and negative parity (panel (b)) states for nuclei pairs 
$^{85}$Se-$^{85}$Br, $^{93}$Sr-$^{93}$Y, $^{92}$Y-$^{92}$Zr, $^{99}$Zr-$^{99}$Nb, $^{102}$Nb-$^{102}$Mo, $^{103}$Mo-$^{103}$Tc, $^{104}$Tc-$^{104}$Ru, and $^{106}$Rh-$^{106}$Pd. All of these nuclei contribute significantly 
to the reactor antineutrino spectra.
The x-axis represents the energy differences between the initial and final states ($E_x = E_f-E_i$) involved in each transition.
The dashed line indicates $\delta_{LS}^{j_fj_i}=-1/2$. 
As discussed in the text, values of $\delta_{LS}^{j_fj_i}$ that deviate significantly from $-1/2$ only occur for transitions
for which matrix elements of the GT operator are quite small.}
\end{figure}

\begin{figure}
\includegraphics[angle=0,width=12 cm]{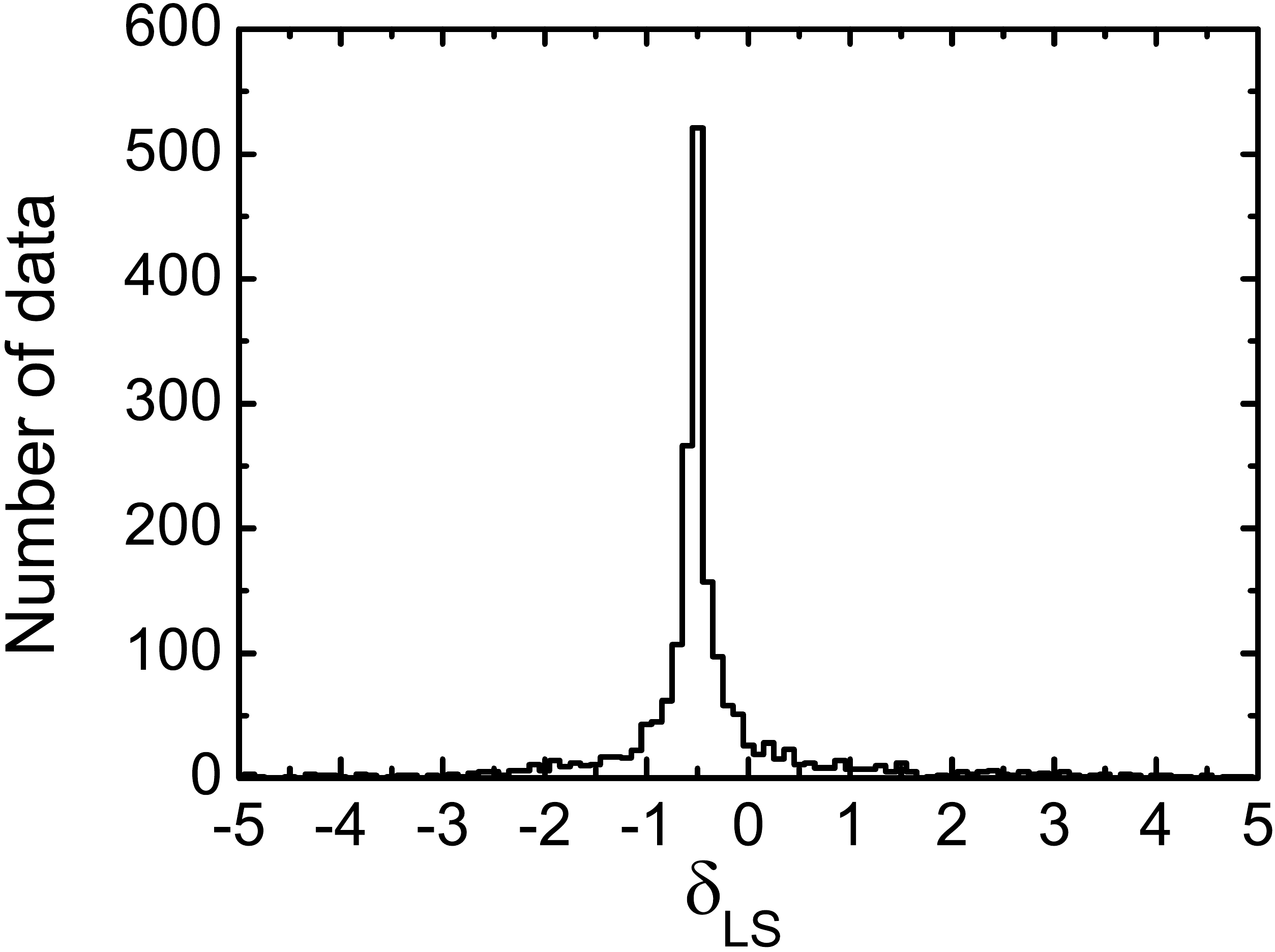}
\caption[T]{\label{fig:delLS-data}
(Color online)  The frequency distribution for $\delta_{LS}^{j_fj_i}$ in Fig.~\ref{fig:delLS}. Bin widths are 0.1. }
\end{figure}
We also examined the allowed beta-decay between the fission fragments within the nuclear many-body shell model.
We first note that many shell model truncation schemes that we examined, containing the most important contribution to the beta-decay transitions of interest and that are
often adopted for the $A\sim100$ and $A\sim130-150$ regions 
of the nuclear mass table,
 automatically result in $\delta_{LS} ^{j_fj_i} =-1/2$.
This is because the only allowed $1^+$ transitions in such schemes involve transitions between
 spin-orbit partners, and, as shown in Sec. IIa, transitions between spin-orbit partners can only yield 
$\delta_{LS}^{j_fj_i}=-1/2$.
For example, 
 the {\it jj45pna} interaction~\cite{mhj95} suitable to the $A\sim 100$ region, which is built on a $^{78}$Ni core, involves the $\pi 1f5/2$, $\pi 2p$ and $\pi 1g9/2$ proton orbitals, and $ \nu 1g7/2$, $ \nu 2d$, $ \nu 3s$ and  $ \nu 1h11/2$ neutron orbitals. 
Similarly, for nuclei in the $A\sim 130-150$ mass region, the {\it jj56} model space for use with the {\it j56cdb} interaction~\cite{cdbonn-132sn},  is built on $^{132}$Sn core,
and involves the $\pi 1g7/2$, $\pi 2d$, $\pi 3s$ and $\pi 1h11/2$ proton orbitals, and the
 $\nu 1h9/2$ $\nu 2f$, $\nu 3p$  $\nu 3s$ and  $\nu 1i13/2$ neutron orbitals.  

Going beyond model spaces of this size for the heavier mass fragments ($A\sim 130$) was not possible in the present study.
One of the issues is the lack of a reliable truncation schemes 
with corresponding effective two-body nucleon-nucleon interactions.
Thus, by the choice of model space,  none of our calculations predicted values for $\delta_{LS}^{j_fj_i}$ different from $-1/2$ for the $A\sim 130$ fragments.
 
For the $A\sim 100$ region, it was easier to handle larger model spaces, and we used {\it glekpn} interaction and model space, 
as described in Ref.~\cite{glek}.
This model space includes five proton orbits ($\pi 1f_{7/2}$, $\pi 1f_{5/2}$, $\pi 2p_{3/2}$, $\pi 2p_{1/2}$ and  $\pi 1g_{9/2}$) and five neutron orbits  ($ \nu 1g_{9/2}$, $\nu 1g_{7/2}$, $\nu 2d_{5/2}$, $\nu 2d_{3/2}$ and  $\nu 3s_{1/2}$). 
This interaction is suitable for describing nuclei with $20 < Z < 50$ and $40 < N < 70$. 
In addition, it has been successfully applied \cite{glek} to describe the first forbidden decay of a dominant 
contribution to fission antineutrino spectra, namely $^{96}$Y.
To calculate the structure of the $A\sim 100$ fission fragments within this model space, 
we used the shell model code {\sc KSHELL}~\cite{kshell}. 
This code corrects for spurious center-of-mass excitations by adding the center-of-mass Hamiltonian, 
multiplied by a large positive coefficient, to the nuclear Hamiltonian, thus pushing spurious states to very high 
excitation energies \cite{com}.
Studying the broad set of nuclei in the $A\sim 100$ regions shown in Fig.~1, including excited states, made it necessary for us to truncate the 
{\it glekpn} model space. This mainly involved freezing the proton configurations. 
For $Z<40$ nuclei, we took the  $\pi 1f_{7/2}$ shell to be full, and only 
allowed two-particle excitations from $\pi 1f_{5/2}$, $\pi 2p_{3/2}$, and $\pi 2p_{1/2}$ shells into  $\pi 1g_{9/2}$; For $Z>40$ nuclei, we froze $\pi 1f2p$ shell for positive parity states, 
and allowed one-particle excitations 
(from the $\pi 2p_{3/2}$ or $\pi 2p_{1/2}$ shells) for negative parity states. 
For the neutron configurations, we allowed one-particle excitations from $ \nu 1g_{9/2}$ shell, and two-particles excitations
 from $ \nu 2d_{5/2}$ shell, at most.  
The $ \nu 1g_{9/2}$ and $ \nu 2d_{5/2}$ orbitals have lower single-particle energies than other neutron orbitals in the {\it glekpn} interaction, which is consistent with the HFB-LN calculations shown in Fig.~\ref{fig:A100}.
 
We calculated the beta decays for the allowed transitions to the lowest 20 positive parity states or lowest 20 negative parity states, depending on the parity of the parent nucleus.
The energy distribution of the resulting value for the ratio $\delta_{LS}^{j_fj_i}$ for eight specific nuclei that contribute significantly to the fission antineutrino spectra is shown in Fig.~\ref{fig:delLS}. 
In this figure, 1947 predicted fission fragment transitions are shown, 
while there were an additional 43 
calculated transitions found to lie outside the range of the figure, $-10.0$ to $ 10.0$. 
These outliers arose mostly because of the poor numerical precision possible for very weak
 transition matrix elements,  $10^{-4}$ or smaller.
Results are shown in the figure for positive-to-positive parity and negative-to-negative parity transitions
separately. The model space for protons, which includes the $\pi 1g_{9/2}$ orbit, contains orbitals with different parities,  while the neutron orbits are all positive parity. 
Thus, the different parity states are produced by configuration mixing between the protons. 
However, as can seen from the figure, the distribution of $\delta_{LS}^{j_fj_i}$ is very similar for the two parity cases.

We further collect the data for  $\delta_{LS}^{j_fj_i}$ 
in Fig.~\ref{fig:delLS-data}, where we show the frequency of occurrence of a given value of $\delta_{LS}^{j_fj_i}$.  
Here there are 1897 data points for the range $-5.0\sim 5.0$, which covers $97.43\%$ of the transitions examined.
 If we average over all the data in Fig.~\ref{fig:delLS-data}, we find, 
\begin{equation}
\delta_{LS}^{j_fj_i} =-0.42\pm 0.99\, ,
\end{equation}
where the error is estimated from the standard deviation. 

We examined the transitions where $\delta_{LS}^{j_fj_i}$ deviated significantly from a value of -1/2 in detail, 
and found that this generally occurred only when matrix elements of $\vec{\Sigma}$ were weak.
Since $\delta_{LS}^{j_fj_i}$ is the ratio of the matrix elements of orbital and spin currents, 
the absolute value of $\delta_{LS}^{j_fj_i}$ can become large and/or  quite uncertain 
whenever the matrix elements  of  $ \la J_f ||\vec{\Sigma} ||J_i\ra$ is less than $10^{-3}$. 
This issue is the dominant effect determining the standard deviation and quoted error for $\delta_{LS}^{j_fj_i}$.

\subsection{Shell-model results for $\delta_{LS}^{j_fj_i}$ for a broader set of nuclei}
\begin{figure}
\includegraphics[angle=0,width=12 cm]{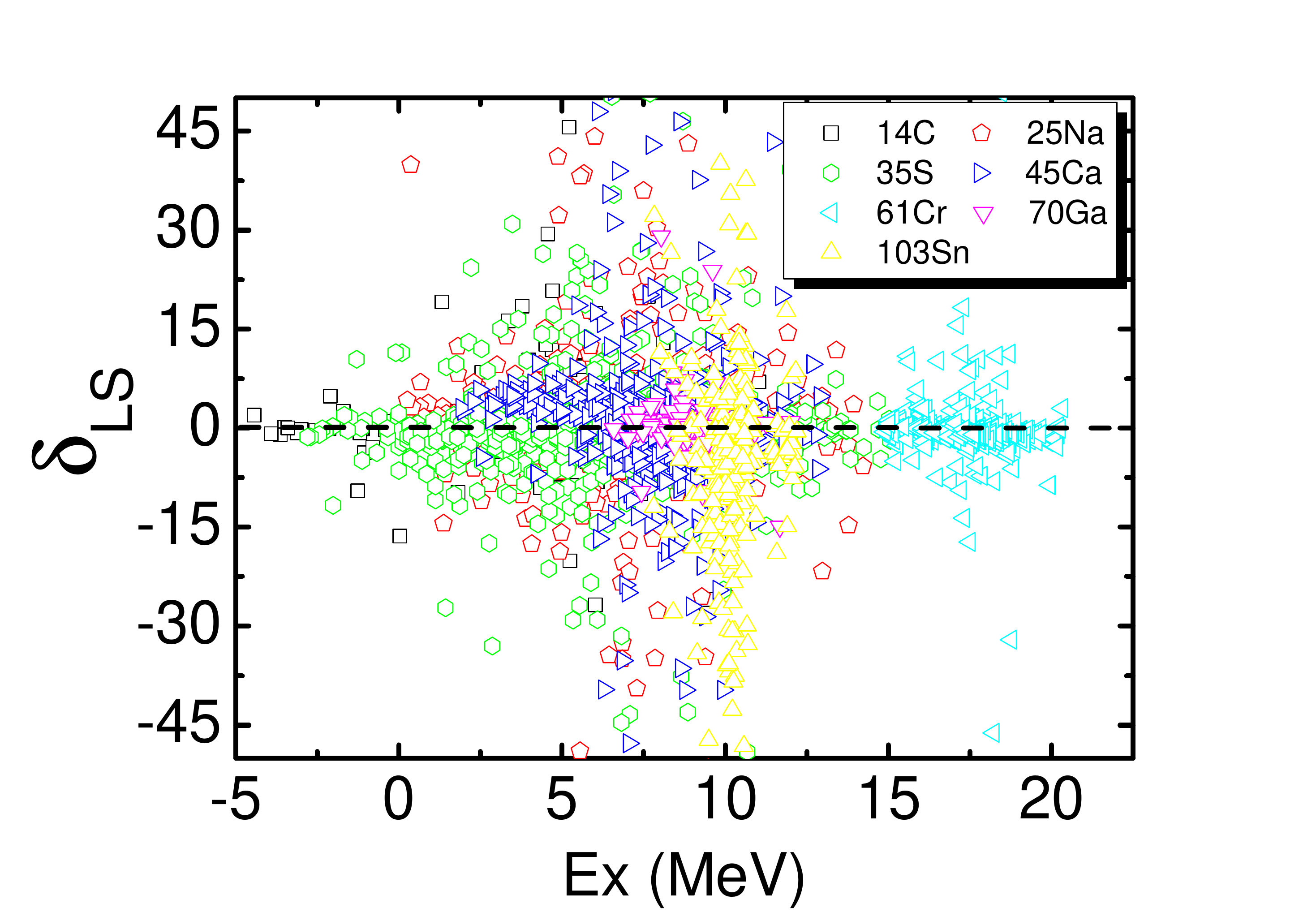}
\caption[T]{\label{fig:more}
(Color online) The set of $\delta_{LS}^{j_fj_i}$ values calculated for the broader set of nuclei with mass range 14-103,
 not including fission fragments, for which the neutrons and protons responsible for the beta-decay transitions are in the same shell.
Typically, only one major shell and 40 lowest states of each daughter nucleus were included. 
In the case of $^{14}$C-$^{14}$N, a multi-shell calculation was carried out using the
 PSDMWK (0+2+4 $\hbar \omega$ truncation)  interaction~\cite{mwk} and Lawson's prescription of center of mass (C.M.) correction (a factor of 10 times the C.M. Hamiltonian)~\cite{com}.
The pairs $^{25}$Na-$^{25}$Mg and $^{35}$S-$^{35}$Cl were studied using the USDB interaction~\cite{usdb},  $^{45}$Ca-$^{45}$Sc, $^{61}$Cr-$^{61}$Mn and $^{70}$Ga-$^{70}$Ge with GXPF1A interaction~\cite{gxpf1a}, and $^{103}$Sn-$^{103}$Sb with {\it jj55pna} interaction~\cite{cdbonn-132sn}. The dashed line is for  $\delta_{LS}^{j_fj_i}=0$.}
\end{figure}

To obtain a more physical understanding of the nuclear structure issues determining the magnitude of the ratio 
$\delta_{LS}^{j_fj_i}$, we studied a set of nuclei that do not contribute to the fission antineutrino spectra but
 that span a much broader range of nuclear masses.
The results of this study, involving nuclei with mass ranging from 14 to 103,  are summarized in Fig.~\ref{fig:more}.
For this chosen set of nuclei, the neutron and proton number do not differ significantly, which allow many ways of forming
 allowed GT transitions. In particular, it allows a study of
 transitions between orbitals of the same $j$ quantum number.
For all of these nuclei, except $^{14}$C, we restricted the shell model calculation to one major shell.
The figure displays the value of $\delta_{LS}^{j_fj_i}$  for  3633 transitions, and the data are
seen to vary in the range of  $-50.0$ to $+50.0$.  
As can be seen from Fig. 7, the corresponding values for $\delta_{LS}^{j_fj_i}$ exhibit a very broad distribution, being significant for the entire range from
 $-10.0$ to $+10.0$. This situation is in sharp contrast to that for the fission fragments contributing to antineutrino spectra,
where the distribution is much narrower, Fig. 5. 

The broad and somewhat random pattern for $\delta_{LS}^{j_fj_i}$ seen 
for nuclei
 in which the neutrons and protons making up the
beta-decay transitions are in the same shell is not completely surprising, 
given the analytical results of SPME for  $\delta_{LS}^{j_fj_i}$ in Eq.~(\ref{delta+-}) when $j_i=j_f$. 
For the general case involving configuration mixing and transitions between neutron and proton orbitals of the same $j$, there is no simple approximation for the quantity $\delta_{LS}^{j_fj_i}$,
and a nuclear structure independent 
analytic expression for the orbital contribution to weak magnetism does not exist. 

However, when large spin-orbit splitting are involved, and the neutron Fermi surface is very different from the proton one, 
the ratio $\delta_{LS}^{j_fj_i}$ tends to be dominated by  values close to -1/2. 
The current work suggests that this situation describes well the strong GT transitions for both the
A$\sim$100 and A$>$130 fission fragments,
because those beta-decays mainly involve transitions between spin-orbit partners, $\pi g9/2$ to $\nu g7/2$ and $\pi h11/2$ to $\nu h9/2$, respectively.
However, for weak transitions, particularly small branches to excited states of the daughter nucleus, matrix elements of $\vec{\Sigma}$ can be suppressed and the ratio of the matrix elements of the orbital to spin current becomes difficult to predict.

\begin{figure}
\includegraphics[angle=0,width=12 cm]{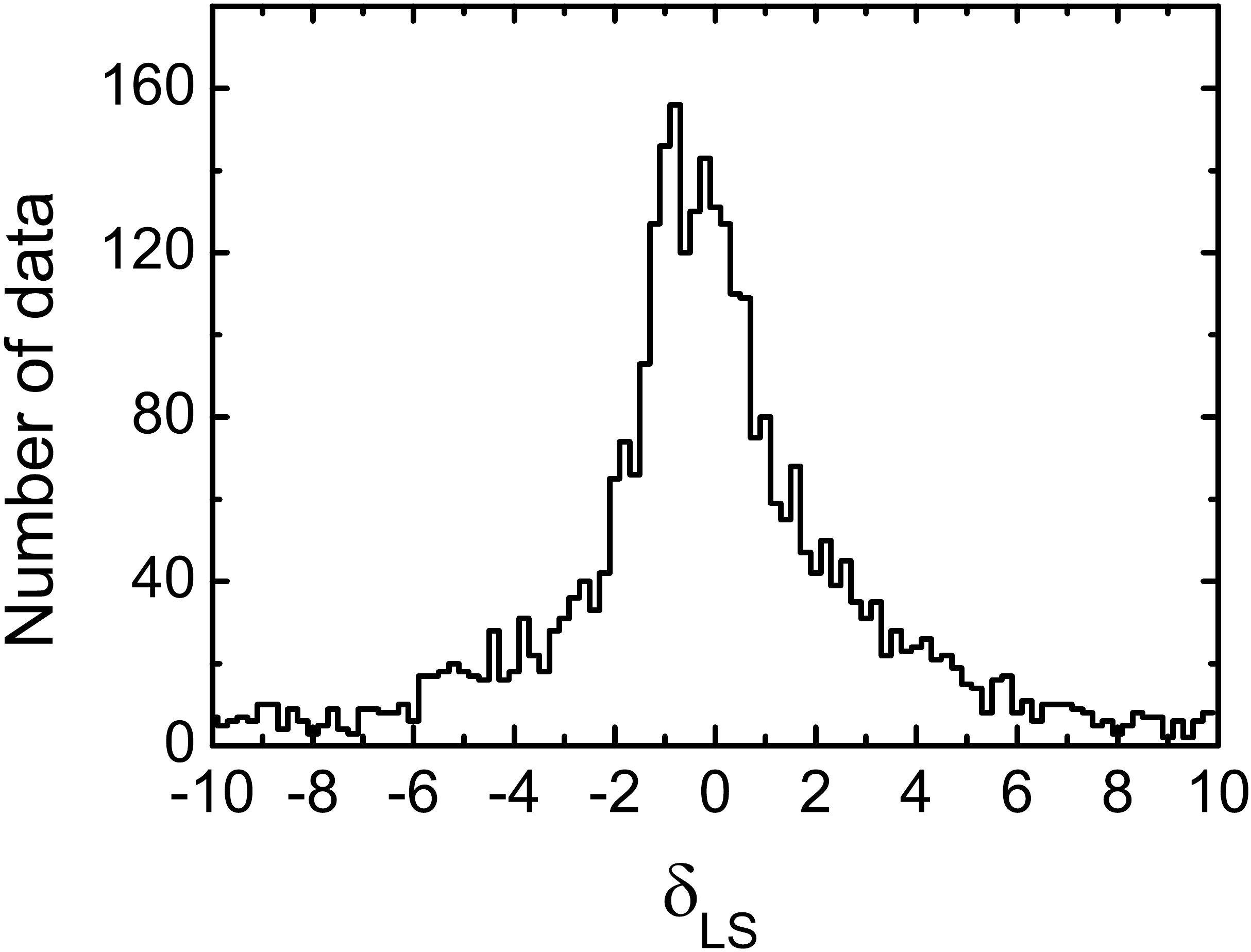}
\caption[T]{\label{fig:more-data}
(Color online)  The frequency distribution for $\delta_{LS}^{j_fj_i}$ corresponding to the nuclei in Fig.~\ref{fig:more}. Bin widths are 0.2. The distribution is much broader than that for fission fragments, Fig. 5, because many of the transition involve
orbitals with $j_i=j_f$ in the same shell. }
\end{figure}


\section{The effect on the uncertainty in one-body WM correction on fission antineutrino spectra}
Our nuclear structure calculations suggest that the value of $\delta_{LS}$ is close to $-1/2$, with a one standard deviation value 
of $\pm1$. In this section, we examine the effect of this uncertainty on the one-body weak magnetism correction 
to allowed beta-decay.
In Fig. 8, we show the ratio of the spectrum for a single beta-decay with different values of $\delta_{LS}^{j_fj_i}$ to that with $\delta_{LS}^{j_fj_i}=-1/2$. In this example the transition  was assumed to be a pure GT one, with an end-point energy of $E_0=6.0+m_ec^2$ MeV.
For all values of $\delta_{LS}^{j_fj_i}$ the spectra are normalized to unity. 
As expected, in all cases, the change in $\delta_{LS}^{j_fj_i}$ leads to a linear change in the shape of the spectrum, which crosses unity at $E_0/2$. 
This change is quite small, being $<2\%$ even when $\delta_{LS}^{j_fj_i}$ is taken to deviate from the mean
 more than three standard deviations.

In Fig. 9, we show the situation for the full aggregate antineutrino spectrum for $^{235}$U thermal fission, 
where the beta-decay end-point energies, branching ratios and the fission yields are taken from ENDF/B-VII.1.
Again, even for very large ($>3$) standard deviations from the predicted mean value of $\delta_{LS}^{j_fj_i}$, 
the change in the shape of the spectrum is less than 2\% at all energy of interest.
If we restrict the uncertainty to two standard deviations, the uncertainty is less than 1\%.
 The narrow distribution for $\delta_{LS}^{j_fj_i}$ in the case of fission fragments, Fig. 5, suggest that the uncertainty
in the one-body weak magnetism contribution to fission antineutrino spectra is closer to $\sim 0.5\%$.
\begin{figure}
\includegraphics[width=10cm]{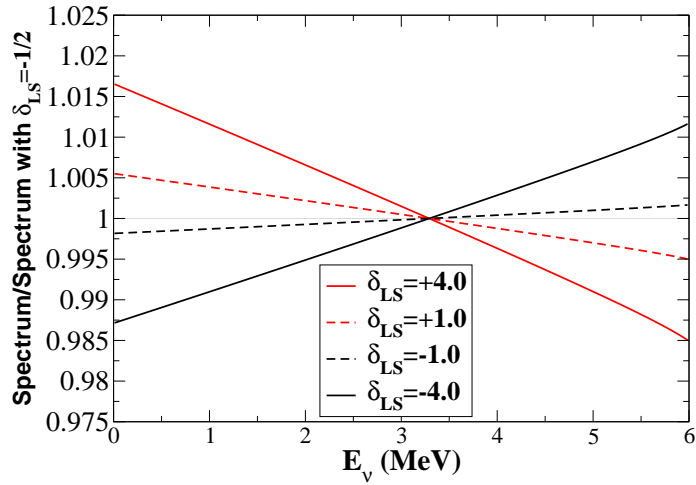}
\caption{The effect of different values of $\delta_{LS}$ on a single transition of endpoint energy 6 MeV.
The figure shows the ratio of spectrum calculated with different values of $\delta_{LS}$ to that with $\delta_{LS}=-1/2$.
A value of $\delta_{LS}=\pm 4.0$, which is more than three standard deviations from the mean, changes the spectrum shape by less than 2\% at any energy of interest. If we restrict the uncertainty to two standard deviations, the uncertainty is less than 1\%.}
\end{figure}
\begin{figure}
\includegraphics[width=10cm]{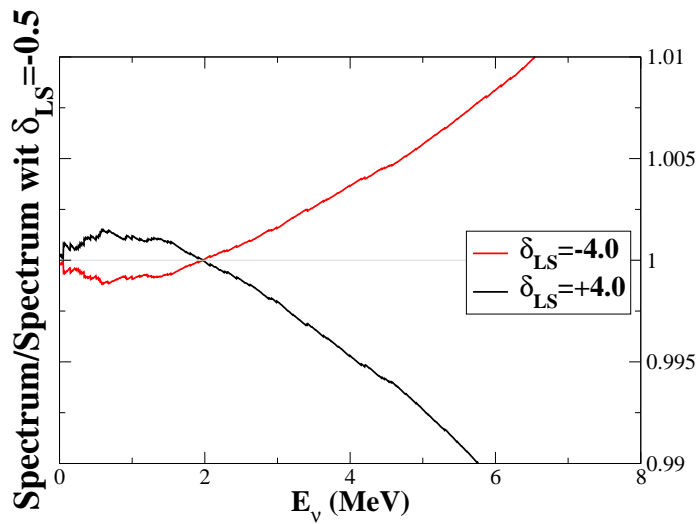}
\caption{The same as Fig. 8, but for the entire aggregate thermal fission antineutrino spectrum for $^{235}$U} 
\end{figure}
\section{summary and conclusion}
The weak magnetism correction to nuclear beta-decay involves three components, resulting from 
the spin, orbital and meson-exchange terms in the magnetic dipole operator.
We have examined the often invoked approximation for the orbital contribution. This approximation assumes that the orbital contribution is proportional to the spin contrition, and that 
$\la J_f ||\vec{\Lambda} ||J_i \ra =-\frac{1}{2}\la J_f ||\vec{\Sigma} ||J_i \ra$.
For a general beta-decay transition, this assumption is not found to be good 
and the one-body weak magnetism corrections require detailed nuclear structure calculations.
However, in the case of the fission fragments that dominate fission antineutrino spectra the 
assumption is found to be a good approximation, and typically introduces less than a 1\% uncertainty in the fission antineutrino spectra. This is because
transitions between the fission fragments of interest are dominated by spin-orbit pairs, in which case 
$\delta_{LS}^{j_fj_i}=-1/2$. 
Contributions to weak magnetism from meson-exchange currents have not been examined and require additional study.

\acknowledgments{We thank J.L. Friar for for very detailed and helpful discussions. 
X.B. Wang wishes to thank the National Natural Science Foundation of China under Grants No. 11505056 and No. 11605054 and China Scholarship Council (201508330016) for supporting his research. A. C. Hayes thanks the Los Alamos National Laboratory LDRD program. This work was partially supported under the U.S. Department of Energy FIRE Topical Collaboration in Nuclear Theory.}


\begin{thebibliography}{99}

\bibitem{mueller}A. Mueller, D. Lhuillier, M. Fallot, A. Letourneau, S. Cormon, M. Fechner,
L. Giot, T. Lasserre, J. Martino, G. Mention, A Porta, and  F. Yermia, Phys. Rev. C {\bf 83}, 054615 (2011).
\bibitem{huber} P. Huber, Phys. Rev. C {\bf 84}, 024617 (2011). 
\bibitem{Schreckenbach} F. Von Feilitzsch, A. Hahn, and K. Schreckenbach, Phys. Lett. {\bf B118}, 162 (1982);
K. Schreckenbach, G. Colvin, W. Gelletly,  and F. Von Feilitzsch, Phys. Lett. {\bf B160} 325 (1985);
A. Hahn, K. Schreckenbach, G. Colvin, B. Krusche, W. Gelletly, Phys. Lett. {\bf B218}, 365 (1989). 
\bibitem{vogel-1} P. Vogel, G. K. Schenter, F. M. Mann, and R. E. Schenter, Phys. Rev. C {\bf 24}, 1543 (1981).
\bibitem{anomaly}G. Mention, {\it et al.}, Phys. Rev. D 83 073006 (2011).
\bibitem{sirlin}A. Sirlin, Phys. Rev. {\bf 164}, 1767 (1967).
\bibitem{sirlin-new} A. Sirlin, arXiv:1105.2842v2 [hep-ph]; (2011).
\bibitem{Hayes14} A. C. Hayes et al., Phys. Rev. Lett. {\bf 112}, 202501 (2014).
\bibitem{hayes2}A.C. Hayes,  J. L. Friar, G. T. Garvey, Duligur Ibeling, Gerard Jungman, T. Kawano, Robert W. Mills, Phys. Rev. {\bf D} 92, 033015 (2015).
\bibitem{bnl}A. A. Sonzogni, T. D. Johnson, and E. A. McCutchan, Phys. Rev. C 91, 011301 (2015).
\bibitem{holstein} B. R. Holstein,  Phys.\ Rev.\ C  {\bf 9}, 1742 (1974). 
\bibitem{zemach16}  X. B. Wang, J. L. Friar, and A. C. Hayes, Phys. Rev. C {\bf 94}, 034314 (2016).
\bibitem{hayes-vogel} Anna C. Hayes and Petr Vogel, Annual Review of Nuclear and Particle Science {\bf 66} 219 (2016). 
\bibitem{Edmonds60}A. R. Edmonds, Angular Momentum in Quantum Mechanics, Princeton University Press, Princeton, 1960.
\bibitem{Cha1998}  E. Chabanat, P. Bonche, P. Haensel, J. Meyer, and R. Schaeffer, {\it Nucl. Phys.} {\bf A635}, 231 (1998).
\bibitem{lipkin} H.J. Lipkin, Ann. of Phys. 9, 272 (1960).
\bibitem{hfodd} N. Schunck, J. Dobaczewski, J. McDonnell, W. Satu{\l}a, J.A. Sheikh, A. Staszczak, M. Stoitsov, P. Toivanen, Comput. Phys. Commun. {\bf 183}, 166 (2012).
\bibitem{NPA1984}  J. Dobaczewski, H. Flocard and J. Treiner,  {\it Nucl. Phys.} {\bf A 422}, 103 (1984).
\bibitem{mhj95}M. Hjorth-Jensen, T. T. S. Kuo, and E. Osnes, Phys. Rep. {\bf 261}, 125 (1995).
\bibitem{cdbonn-132sn} B. A. Brown, N. J. Stone, J. R. Stone, I. S. Towner, and M. Hjorth-Jensen, Phys. Rev. C {\bf 71}, 044317 (2005).
\bibitem{glek}H. Mach, E. K. Warburton, R. L. Gill, R. F. Casten, J. A. Becker, B. A. Brown, and J. A. Winger, Phys. Rev. C {\bf 41}, 226 (1990)
\bibitem{kshell} N. Shimizu, arXiv:1310.5431 (2013).
\bibitem{com}  D.H. Gloeckner, R.D. Lawson, Phys. Lett. {\bf B 53}, 313 (1974).
\bibitem{mwk}E. K. Warburton and B. A. Brown, Phys. Rev. C  {\bf 46}, 923 (1992); E. K. Warburton, B. A. Brown, and D. J. Millener, Phys.Lett.  {\bf B293}, 7 (1992).
\bibitem{usdb}B. Alex Brown and W. A. Richter, Phys. Rev. C  {\bf74}, 034315 (2006).
\bibitem{gxpf1a}M. Honma, T. Otsuka, B.A. Brown, and T. Mizusaki, Eur. Phys. J. A {\bf 25}, s01, 499–502 (2005).
\end{thebibliography}
\end{document}